\documentclass[journal=nalefd,manuscript=letter]{achemso}

\usepackage{amsmath,graphicx,latexsym,times,color}
\usepackage{setspace}
\usepackage{hyperref}
\usepackage{array}
\usepackage{physics}
\usepackage{color}
\usepackage[T1]{fontenc}
\usepackage[version=3]{mhchem} % Formula subscripts using \ce{}
\usepackage{booktabs}
\usepackage{amssymb}
\usepackage{xcolor}
\usepackage{makecell} % Add in preamble
\usepackage{lettrine}
\usepackage[dvipsnames]{xcolor} 
\usepackage{empheq} %% loads mathtools, which loads amsmath
\author{Megha Arya}
\affiliation{CEMES, Universit\'e de Toulouse, CNRS, 29 rue Jeanne Marvig, F-31055 Toulouse, France}

\author{Moritz A. Goerzen}
\affiliation{CEMES, Universit\'e de Toulouse, CNRS, 29 rue Jeanne Marvig, F-31055 Toulouse, France}

\author{Lionel Calmels}
\affiliation{CEMES, Universit\'e de Toulouse, CNRS, 29 rue Jeanne Marvig, F-31055 Toulouse, France}

\author{Rémi Arras}
\affiliation{CEMES, Universit\'e de Toulouse, CNRS, 29 rue Jeanne Marvig, F-31055 Toulouse, France}

\author{Soumyajyoti Haldar}
\affiliation{Institute of Theoretical Physics and Astrophysics, University of Kiel, Leibnizstrasse 15, 24098 Kiel, Germany}
 
\author{Stefan Heinze}
\affiliation{Institute of Theoretical Physics and Astrophysics, University of Kiel, Leibnizstrasse 15, 24098 Kiel, Germany}
\alsoaffiliation{Kiel Nano, Surface, and Interface Science (KiNSIS), University of Kiel, 24118 Kiel, Germany}

 \author{Dongzhe Li}
\email{dongzhe.li@cemes.fr}
\affiliation{CEMES, Universit\'e de Toulouse, CNRS, 29 rue Jeanne Marvig, F-31055 Toulouse, France}

\title[\texttt{achemso} demonstration]
{A new skyrmion topological transition driven by higher-order exchange interactions in Janus MnSeTe}

\newcommand{\V}[1]{\ensuremath{\mathbf{#1}}} %Vector

\let\oldtimes\times  % Make the times "x" use less spacing
\renewcommand\times{{\oldtimes}}

\definecolor{pacificb}{HTML}{1CA9C9}

\renewcommand{\vec}[1]{\mathbf{#1}}


\usepackage[table]{xcolor}
\usepackage{booktabs}
\definecolor{lightgray}{gray}{0.9}

\begin{document}
	

	%\begin{abstract}
\noindent{\bf{\textcolor{red}{ABSTRACT:}}} Two-dimensional (2D) van der Waals magnets offer a promising platform for pushing skyrmion technology to the single-layer limit with high tunability. While Dzyaloshinskii–Moriya interaction (DMI) is often recognized as central to skyrmion formation, their stability, collapse, and topological transition in 2D materials remain largely unexplored. In particular, the effect of higher-order exchange interactions (HOI) on these phenomena is unknown. 
Here, using first-principles calculations and atomistic spin simulations, we report a new topological transition generated by HOI, which we term `ferric transition', in single-layer MnSeTe. Surprisingly, skyrmion stability and collapse remain largely unaffected by HOI due to the dominant role of DMI near the saddle point, whereas the Bloch point is strongly modified, giving rise to this novel transition. This mechanism is fundamentally distinct from the well-known radial and chimera transitions. Moreover, we predict that Janus MnSeTe exhibits remarkably high skyrmion energy barriers due to its strong DMI, among the highest reported for intrinsic 2D magnets. Our findings unveil an unexpected role of HOI in skyrmion topological transitions and establish Janus MnSeTe as a robust platform for 2D skyrmionics.
~\\
~\\
		\noindent{\bf{\textcolor{red}{KEYWORDS:}}} Magnetic skyrmions, Dzyaloshinskii–Moriya interaction, Higher-order exchange interactions, ferric topological transition, single-layer MnSeTe 		\\
%	\end{abstract}

\noindent{\bf{\textcolor{red}{TABLE OF CONTENT:}}}

\begin{figure*}[h]%---------------------------------------------------------------------------------------------------------------------
    \centering
    \includegraphics[width=0.7\linewidth]{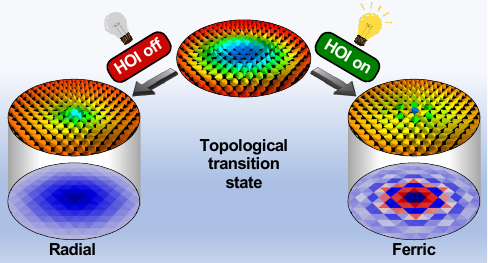}
\end{figure*}%--------------------------------------------------------------------------------------------------------------------------
    
\lettrine[lines=2, lhang=0.05, loversize=0.15]{\textcolor{red}{M}}{\textnormal{agnetic}} skyrmions are vortex-like localized spin textures with non-trivial topology that have raised high hopes for next-generation spintronics devices \cite{fert2017magnetic, everschor2018perspective, luo2018reconfigurable, li2021magnetic, psaroudaki2021skyrmion}. During the last decade, most of the efforts were devoted to studying skyrmions in bulk 
crystals \cite{mühlbauer2009skyrmion, yu2011near, yang2012strong}, ultrathin films with interfaces \cite{heinze2011spontaneous, romming2013writing, dupe2014tailoring, meyer2019isolated}, and multilayers \cite{moreau2016additive, soumyanarayanan2017tunable, raju2019evolution}. 
More recently, the study of skyrmions has progressed to two-dimensional (2D) van der Waals (vdW) magnets \cite{han2019topological,ding2019observation,schmitt2022skyrmionic,zhang20242d}.
Stabilizing skyrmions in 2D vdW magnets offers several potential advantages. These include avoiding pinning by defects due to high-quality vdW interfaces, the possibility of intrinsic Dzyaloshinskii-Moriya interaction (DMI) within a single layer, and easy control of magnetism via external stimuli such as strain \cite{li2022strain},  magnetic proximity \cite{wu2020neel,Dongzhe_prb2023}, electric field \cite{sun2020controlling,Huang2022}, Moiré engineering \cite{Tong2018}, or light \cite{khela2023laser}. An all-electrical scheme based on tunnel junctions was recently proposed to detect skyrmions in 2D magnets \cite{li2024proposal}.

A major challenge in skyrmion-based device applications is that the activation energy for switching a magnetic bit is usually bounded by energy barriers, which ensure the thermal stability of the stored information. It is known that skyrmions are strictly topologically protected in the continuum, meaning a skyrmion cannot be continuously unwound into a uniform ferromagnetic (FM) state. Instead it is often assumed that the skyrmion annihilates by shrinking to a singular point so that its energy reaches the topological Belavin-Polyakov limit \cite{polyakov22metastable}, which consequently gives rise to a finite energy barrier \cite{bernand2022micromagnetic, buttner2018theory}. However, this model breaks down for magnetic textures spanning only a few lattice constants \cite{leonov2016properties} and thus cannot capture more complex annihilation mechanisms that have been reported numerically and experimentally \cite{muckel2021experimental} for skyrmions at the nanoscale.
Therefore, the unbiased investigation of skyrmion collapse and topological transition mechanisms is crucial for enhancing the stability of skyrmions with technologically relevant sizes \cite{muckel2021experimental,li2022strain,li2024stability}.

According to the extended Heisenberg model, the existence of skyrmions is ascribed to a delicate balance between various magnetic interactions, including Heisenberg exchange, DMI, magnetocrystalline anisotropy, and Zeeman interaction. Among these interactions, the DMI, which favors a canting of the spins of adjacent magnetic atoms, is often recognized
as the key ingredient in assigning a unique rotational sense to skyrmions. DMI originates from spin-orbit coupling (SOC) and broken inversion symmetry. However, most 2D magnets are centrosymmetric, and several strategies have been proposed to achieve DMI in these materials by breaking inversion symmetry \cite{liang2020very,li2023topological,Dongzhe_PRB2024}. 

While widely used, the extended Heisenberg model is a localized spin model, limiting its applicability to 3$d$ transition metals due to their itinerant nature of magnetism. The alternative Hubbard model describes electrons on a lattice by their hopping between sites and mutual Coulomb repulsion. Note that the conventional Heisenberg exchange can be obtained in second-order perturbation theory from the $S=1/2$ Hubbard model. 
If one includes higher-order terms in the perturbative expansion, new interactions termed higher-order exchange interactions (HOI) arise. In particular, the 4-spin 2-site (biquadratic) and the 4-spin 4-site interactions occur in fourth-order perturbative expansion for the $S=1/2$ Hubbard model~\cite{takahashi1977half, macdonald1988t}. It has recently been demonstrated that a 4-spin 3-site interaction also arises in $S\geq1$ systems ~\cite{hoffmann2020systematic}, appropriate for 
describing crystals in which the magnetic moments are of
the order of 2 to 3~$\mu_\textrm{B}$. 
It has been demonstrated that these HOI terms 
can play a crucial role for the magnetic properties of ultrathin films and 2D vdW magnets
~\cite{kurz2001three, kronlein2018magnetic, heinze2011spontaneous, spethmann2020discovery,kartsev2020biquadratic, paul2020role,haldar2021,
gutzeit2022nano,
xu2022assembling,pan2024chiral, li2023topological}. 

Janus monolayer transition-metal dichalcogenides, with their inversion asymmetry and strong SOC, 
have been proposed
as a promising platform for skyrmions. Over the last five years, Janus monolayers have been extensively studied by different groups~\cite{liang2020very, yuan2020intrinsic,changsong2020,changsheng2022,zhou2024emergence}.
Two magnetic Janus monolayers have recently been synthesized, thus opening further opportunities for experimental investigation of topological magnetism in these materials~\cite{xu2025unusual, nie2024regulated}. It is worth noting that previous studies on magnetic Janus only focused on realizing topological spin textures using the standard extended Heisenberg model. However, key aspects such as quantification of skyrmion stability, collapse, and topological transition -- all crucial for device applications -- have so far not been reported. In particular, the role of HOI in skyrmion collapse and topological transition are also unknown in 2D vdW magnets.

In this Letter, we report a new skyrmion topological transition induced by HOI in single-layer MnSeTe using first-principles density functional theory (DFT)-based calculations and atomistic spin simulations. This new transition mechanism is termed ferric, which is fundamentally different from the conventional radial \cite{von2017enhanced,li2022strain,goerzen2023lifetime} and chimera \cite{Levente2017,muckel2021experimental} transitions reported in the literature. We explain its origin by using energy-decomposed density maps with HOI.
Additionally, due to the strong DMI originating from the combined effects of broken inversion symmetry and SOC, Néel-type skyrmions are formed at zero magnetic field. Remarkably, we find an energy barrier exceeding 330~meV, which is among the highest reported for intrinsic 2D vdW magnets.

\textbf{First-principles atomistic spin model.}
Fig.~\ref{Structure_Constants}a shows the atomic structure of monolayer MnSeTe. The Mn atoms with point group \( C_{3v} \) form a hexagonal lattice, sandwiched by two atomic planes of Se and Te atoms. We find that the lattice constant is 3.68~\text{\AA}, which is in good agreement with previously reported values in Ref.~\cite{liang2020very} (see Sec.~I of Supporting Information (SI) for details).

We describe the magnetic state of the MnSeTe monolayer by a set of classical magnetic moments $\vec{M}_i$ localized on the atomic sites $i$ of a hexagonal lattice and with a dynamics governed by the following Hamiltonian of the beyond-extended Heisenberg model:
\begin{equation}\label{model}
%\begin{split}
H =-\sum_{ij}J_{ij}(\vec{m}_i \cdot \vec{m}_j)-\sum_{ij}\vec{D}_{ij} \cdot(\vec{m}_i \times \vec{m}_j)\\
 - K_u \sum_i (m_i^z)^2 - \sum_i \mu(\V{m}_i \cdot B) + H_{\text{HOI}}
%\end{split}
\end{equation}
where the $z$-axis is perpendicular to the MnSeTe layer and
\begin{equation}
\begin{split}
	\label{eq2}
		H_{\text{HOI}} = & -B_1\sum_{<ij>}{(\V{m}_i \cdot \V{m}_j)}^2 \\
        & -2Y_1\sum_{<ijk>}(\V{m}_i \cdot \V{m}_j)(\V{m}_j \cdot \V{m}_k)\\
		& -K_1\sum_{<ijkl>}[(\V{m}_i \cdot \V{m}_j)(\V{m}_k \cdot \V{m}_l) +(\V{m}_i \cdot \V{m}_l)(\V{m}_j \cdot \V{m}_k)-(\V{m}_i \cdot \V{m}_k)(\V{m}_j \cdot \V{m}_l)]
    \end{split}
\end{equation}
Here $\vec{m}_i = \vec{M}_i/M_i$ and $\V{m}_j$ are normalized magnetic moments at positions $\vec{R}_i$ and $\vec{R}_j$, respectively. The first four magnetic interaction terms in Eq. \ref{model} correspond to Heisenberg exchange, DMI, magnetocrystalline anisotropy energy (MAE), and Zeeman energy under an external magnetic field $B$ respectively, and they are characterized by the constants $J_{ij}$, $\vec{D}_{ij}$, $K_u$, and $B$ in the related terms. The fifth magnetic interaction term, detailed in Eq. \ref{eq2} denotes HOI energies where $B_1$, $Y_1$, and $K_1$ are biquadratic, 4-spin 3-site, and 4-spin 4-site constants, respectively. In our material model, we systematically considered Heisenberg exchange parameters up to the tenth nearest-neighbor (NN), DMI up to the seventh NN (see Fig.~\ref{Structure_Constants}b for the first three DMI vectors), and HOI only for NN spins. The notations $<ij>$, $<ijk>$, etc., in Eq. \ref{eq2} indicate that the summation is restricted to NN spins (see Fig.~\ref{Structure_Constants}c for NN hopping paths considered). 

To determine the magnetic interaction constants in our system, DFT calculations were performed
with the \textsc{fleur} code \cite{fleurv26,kurz2004ab,heide2009describing} based on the full-potential linearized augmented plane wave (FLAPW) formalism (see Sec.~I of SI for methodology and computational details). Spin-spirals characterized by wave vector \textbf{q} are exact solutions of the extended Heisenberg model 
and often denoted as single-\textbf{q} states. There also exist multi-\textbf{q} states, which correspond to superpositions of these spin spirals. The converged energies of single-\textbf{q} states were utilized to determine pair-wise interaction constants ($J_{ij}$, $\vec{D}_{ij}$, and $K_u$), while the converged energy differences between multi-\textbf{q} and their corresponding single-\textbf{q} states were used to determine multi-spin interactions ($B_1$, $Y_1$, and $K_1$).

\begin{table}[tbp]
\centering
\renewcommand{\arraystretch}{1.2}
\begin{tabular}{lcccccccccc}
\toprule
\multicolumn{10}{c}{\textbf{Heisenberg exchange constants (meV/atom)}} \\
\midrule
\rowcolor{lightgray}
$J_1$ ($J_1'$) & $J_2$ ($J_2'$) & $J_3$ ($J_3'$) & $J_4$ & $J_5$ & $J_6$ & $J_7$ & $J_8$ & $J_9$ & $J_{10}$ \\
3.91 (4.95) & 3.43 (4.47) & 2.22 (1.18) & $-0.75$ & $-0.84$ & 0.09 & 0.53 & 0.16 & 0.06 & $-0.01$ \\
\midrule
\multicolumn{10}{c}{\textbf{DMI constants and MAE (meV/atom)}} \\
\midrule
\rowcolor{lightgray}
$D_1$ & $D_2$ & $D_3$ & $D_4$ & $D_5$ & $D_6$ & $D_7$ & $K_u$ & & \\
$-1.04$ & 0.19 & $-0.42$ & 0.01 & 0.20 & $-0.01$ & $-0.03$ & 0.42 & & \\
\midrule
\multicolumn{10}{c}{\textbf{HOI constants (meV/atom)}} \\
\midrule
\rowcolor{lightgray}
$B_1$ & $Y_1$ & $K_1$ & & & & & & &  \\
2.07 & $-1.04$ & $-0.89$ & & & & & & &  \\
\bottomrule
\end{tabular}
\caption{Shell-resolved Heisenberg exchange constants ($J_i$), HOI-modified exchange constants ($J_i'$), DMI constants ($D_i$), MAE constant $K_u$, and NN HOI constants ($B_1$, $Y_1$, $K_1$) for MnSeTe. Positive (negative) $J_i$ indicates FM (AFM) coupling. Positive (negative) $D_i$ favors CW (CCW) spin spirals. $K_u>0$ favors an out-of-plane magnetization.}
\label{table1}
\end{table}

 Firstly, we focus on spin spirals \textbf{q} along the high-symmetry paths ${\overline{\Gamma\mathrm{M}}}$ and ${\overline{\Gamma\mathrm{K}\mathrm{M}}}$ of the Brillouin zone (BZ) (Fig.~\ref{SSstates_dispersion}b and c). The high-symmetry points represent special magnetic states: $\overline{\Gamma}$ corresponds to the FM state, $\overline{\mathrm{K}}$ to the N\'eel-state with an angle of 120$^{\circ}$ between adjacent spins, and $\overline{\mathrm{M}}$ to the row-wise anti-ferromagnetic (RWAFM) state.  A fit of the spin spiral energy dispersions, calculated with and without SOC, to the corresponding terms of the Heisenberg model, respectively, yields the DMI and the exchange energy constants displayed in Table~\ref{table1}. 
Without SOC, the energy dispersion shows a minimum at $\overline{\Gamma},$ which represents the FM state. Note that the chalcogen atoms exhibit small induced magnetic moments, causing deviations from an ideal magnetic monolayer structure. Consequently, the DFT energies at $\overline{\mathrm{M}}$ for spin spirals along different directions are not equivalent. Furthermore, analytical functions derived from the monolayer model show slight deviations when fitted near $\overline{\mathrm{M}}$ due to these deviations. When SOC is turned on, a significant DMI is observed as expected and in agreement with the literature~\cite{liang2020very, yuan2020intrinsic}, and it favors a counterclockwise (CCW) rotational sense, which corresponds to the negative value of energy $\Delta E_{\text{SOC}} (\textbf{q})$. We further confirm that the DMI can be explained by the Fert-Lévy model \cite{Fert1980} since it mainly originates from interfacial nonmagnetic atoms (see Sec.~II of SI). SOC is also responsible for the finite out-of-plane MAE as presented in Table~\ref{table1}. The MAE shifts the energy of spin spirals by $K_u/2$ with respect to the FM state since on average half of the spins are orientated in-plane and half of the spins out-of-plane. The magnetic ground state is found to be the FM state when we include all terms from DFT, i.e., $E_{\text{ss}}(\textbf{q}) + \Delta E_{\text{SOC}}(\textbf{q}) + K_u/2$.

Next, we focus on three prototypical multi-\textbf{q} states\cite{hardrat2009complex,kurz2001three}, shown in Fig.~\ref{SSstates_dispersion}a. Among them, the two \textit{uudd} states are lower in energy, whereas the $3Q$ state is higher in energy compared to the corresponding $1Q$
states studied without SOC. Also, all three are far above the FM state (Fig.~\ref{SSstates_dispersion}b). For the $3Q$ state, there are two possible choices of the corresponding single-\textbf{q} reference state: one at the $\overline{\text{M}}$ point along $\overline{\Gamma \text{M}}$ and another at the $\overline{\text{M}}$ point along $\overline{\Gamma \text{KM}}$. By comparing the magnetic moments of these RWAFM states at both $\overline{\text{M}}$ points with that of the $3Q$ state, we find that the one along $\overline{\Gamma \text{M}}$ provides a closer match. Therefore, we use the RWAFM state along  $\overline{\Gamma \text{M}}$ for further analysis in the main text (note that we have also performed analysis and comparison using the RWAFM state along  $\overline{\Gamma \text{KM}}$ and found that the general conclusions remain unchanged; see Sec.~III of SI). The calculated HOI constants are displayed in Table~\ref{table1}. The biquadratic constant is comparable to the third NN exchange constant, while the 4-spin 3-site and 4-spin 4-site constants have smaller magnitudes and opposite signs relative to the biquadratic constant.

Note that the functional forms of the biquadratic and the 4-spin 3-site interactions for spin spirals resemble those of the first three exchange constants \cite{paul2020role}.
Hence, we cannot separate these exchange and HOI constants by fits and, to avoid double counting, we thus computed corrected exchange constants $J_{1}'$, $J_{2}'$, and $J_{3}'$, as displayed in Table~\ref{table1}.
%?} 
Nonetheless, the exchange interaction remains dominated by the first two NN constants.

\textbf{Atomistic spin simulations.}
We performed atomistic spin simulations using the \textsc{Spinaker} code to explore the properties of isolated skyrmions in Janus MnSeTe (see Sec.~I of SI for methodology and computational details). Our simulations were performed using two distinct parameter sets from Table~\ref{table1}: one utilizing $J_i$ and $D_i$, referred to as "without HOI" simulations, and the other utilizing $J_1'$-$J_3'$, $J_4$-$J_{10}$, $D_i$, as well as $B_1$, $Y_1$, and $K_1$, referred to as "with HOI" simulations. On the other hand, we have used MAE~$= K_u/2$ in our calculations since this quantity, which is sensitive to doping or temperature, can easily be adjusted in experiments~\cite{tan2018hard, wang2020modifications, park2019controlling}. Moreover, as we have checked, the use of different MAE values such as $K_u/5$ or $K_u/10$ does not change the general conclusions of this work.

Using spin dynamics, we observe the emergence of N\'{e}el-type skyrmions
at zero magnetic field, both with and without HOI. To quantify skyrmion size, we employ two definitions that are commonly used in the literature -- one proposed by Bocdanov, \textit{et al.}~\cite{bocdanov1994properties} ($R_{\text{Boc}}$) and another often used in experiments~\cite{wang2018theory}  ($R_{\text{Exp}}$) (see Sec.~IV of SI). These two approaches yield noticeably different size estimates -- for example, at $B = 0$~T without HOI, $R_{\text{Boc}} \approx 1.6 R_{\text{Exp}}$. As the magnetic field increases, the skyrmion size shrinks rapidly according to both definitions, as shown in Fig.~\ref{radii_barrier}a. However, the inclusion of HOI has a negligible effect on the skyrmion size. Without HOI, $R_{\text{Boc}}$ decreases from about 17.1~nm at $B = 0$~T to about 5.0~nm at $B = 0.75$~T; with HOI, it changes slightly from 17.1~nm to 4.9~nm over the same field range. Similarly, $R_{\text{Exp}}$ decreases from about 10.7~nm to 2.8~nm without HOI, and from 10.7~nm to 2.6~nm with HOI, as $B$ increases from 0 to 0.75~T. 

To quantify the thermal stability of skyrmions, we calculate the energy barriers that stabilize skyrmions against collapse, both with and without HOI (see Fig.~\ref{radii_barrier}b). This has been done using the geodesic nudged elastic band (GNEB) method
\cite{bessarab2015method}.
Surprisingly, these skyrmions exhibit high stability with energy barriers exceeding 330~meV at zero magnetic field -- even surpassing those reported for state-of-the-art ultrathin films \cite{von2017enhanced}. The energy barriers exhibit a nonlinear dependence on the magnetic field and remain sufficiently high for $B$ values up to 0.7 T, making them technologically desirable for skyrmionic devices. In the insets of Fig.~\ref{radii_barrier}, we show the difference between the cases with and without HOI. Relatively small deviations are observed for both the radius and energy barriers. In view of previous work on the large enhancement of skyrmion stability due to HOI in ultrathin films \cite{paul2020role}, it is quite unexpected that we find the energy barrier with HOI to be only slightly lower than that obtained neglecting HOI.

To get more insight into the effect of HOI, we plot in Fig.~\ref{MEP}a the minimum energy path (MEP) for the skyrmion collapse into the FM ground state calculated using the GNEB method at $B = 0$~T. The positions of saddle point (SP) and Bloch point (BP) are marked along the MEP for both cases, with and without HOI. The SP is the point of maximum energy w.r.t the initial skyrmion state. The BP is identified by calculating the topological charge, given by
$Q=\frac{1}{4\pi} \int_{\mathbb{R}^2} \V{m} \cdot (\frac{\partial \V{m}}{\partial x} \times \frac{\partial \V{m}}{\partial y})\mathrm{d}x\mathrm{d}y$,  which jumps from $-1$ to $0$ at the BP. Interestingly, we find that the SP and the BP do not coincide in either case.
While this separation of SP and BP is unusual in continuum models, where the zero-size skyrmion \cite{bernand2022micromagnetic, buttner2018theory} resembles both at the same time, there is no law for a discrete spin model stating that the BP has to be the point of highest energy on a reaction coordinate (i.e., SP). Thus, our findings underline the complexity of skyrmion topological transitions, even though, to the best of our knowledge, this separation has not explicitly been addressed in the literature so far. 

From the analysis of spin structures and topological charge densities (see Sec.~V of SI) at different points along MEP (see Fig.~\ref{MEP}b), we find that the SPs are almost identical in both cases. However, surprisingly, the skyrmion topological transitions are very different.
Without HOI, we observe a radially symmetric transition, where the skyrmion shrinks symmetrically until the BP and finally collapses into the FM state. Remarkably, with HOI, a new skyrmion topological transition, which we denote as ferric transition, is observed. We call it ferric because of the quasi-ferrimagnetic (FI) state that emerges as a metastable state in the transition induced by HOI. 
This new transition is fundamentally distinct from both radial \cite{von2017enhanced,li2022strain,goerzen2023lifetime} and chimera \cite{Levente2017,muckel2021experimental} transitions reported in the literature.
The skyrmion transitions into a quasi-FI state wherein opposing topological charge density substructures are formed near the center, leading to partial topological cancellation before finally collapsing into the FM state (see Sec.~VI of SI for additional figures illustrating ferric transition at finite magnetic fields). From site-resolved energy contributions along the MEP, presented in Sec.~VII of SI, it becomes clear that 4-spin 3-site and 4-spin 4-site interactions favor the local FI modulation. Considering the Hamiltonian utilized in this work, the energy difference per site between FM and FI state on the hexagonal lattice is $E_{\text{FI}} - E_{\text{FM}} = 16(Y_1+K_1) + E_{\text{RWAFM}}^{\text{exc}}-E_{\text{FM}}^{\text{exc}}$, where the last two terms are the
Heisenberg exchange energies of the RWAFM and FM state, respectively, leading to the FI being favored by both HOI terms in MnSeTe (See Sec.~VII of SI for details). 
Since the dominating frustrated exchange interaction still enforces a FM ground state, the FI modulation at the BP appears due to $Y_1,K_1<0$ and consequently leads to a reduction in energy compared to the case without HOI, as shown in Fig.~\ref{MEP}a. So, HOI significantly decreases the energies after the SP, leading to very different BP configurations. However, HOI have only a minor effect on the energy barrier, as it is dominated by DMI. As a result, the interplay between DMI and HOI results in almost identical SPs with and without HOI, but their BPs are completely different.

Furthermore, we investigated the energy decomposition of MEP at $B = 0$~T for both cases (see Fig.~\ref{Decomposition}). We find that the energy barriers arise from the competition between various interactions, with DMI strongly favoring the skyrmion state, while exchange and MAE favor the FM state.
Since HOI terms modify the exchange constants, a meaningful comparison would be between the exchange contribution without HOI and the combined exchange (which includes Heisenberg exchange, biquadratic, and 4-spin 3-site interactions) contribution with HOI {\cite{paul2020role}. Interestingly, these contributions are of comparable magnitude at the SP in both cases. Also, the DMI and MAE contributions are of similar magnitudes at the SP with and without HOI. The slight difference is an associated effect of the HOI caused by the changes in relative spin angles during the collapse process. The biquadratic and 4-spin 3-site interactions exhibit nearly compensating contributions. A striking observation is that the 4-spin 4-site interaction acts in a qualitatively different way compared to all other interactions
in agreement with previous work on the role of HOI for skyrmion stability in transition-metal films~\cite{paul2020role}. It remains nearly constant along most of the MEP but exhibits a dip just after the SP. This explains why the transition mechanism is altered, but the energy barriers remain largely unchanged with and without HOI. In addition, we observe that the 4-spin 3-site interaction favors the skyrmion state, whereas the biquadratic and the 4-spin 4-site interaction favor the FM state.  

To conclude, we demonstrate that a fundamental skyrmion transition emerges in the presence of HOI. We term this topological transition as ferric transition and show that it possesses unique characteristics that make it distinct from the conventional radial and chimera transitions. We explain the origin of this new skyrmion topological transition and further identify the relevant HOI terms that cause it. Our predictions can be experimentally tested using, for instance, spin-polarized scanning tunneling microscopy \cite{muckel2021experimental}. Additionally, due to the strong DMI in Janus MnSeTe, we find skyrmion energy barriers exceeding 330 meV at zero field. These values are among the highest reported for intrinsic 2D magnets. Our work demonstrates that taking into account the contribution of HOI is likely to be essential for skyrmion topological transitions.

%#################################################
	
\textbf{Acknowledgement:} This study has been supported through the ANR grant no.~ANR-22-CE24-0019. This work is supported by France 2030 government investment plan managed by the French National Research Agency under grant reference PEPR SPIN – [SPINTHEORY] ANR-22-EXSP-0009. This study has been (partially) supported through the grant NanoX no.~ANR-17-EURE-0009 in the framework of the "Programme des Investissements d’Avenir". S.Ha and S.He gratefully acknowledge financial support from the Deutsche Forschungs Gemeinschaft (DFG, German Research Foundation) through SPP2137 "Skyrmionics" (project no.~462602351) and 
project no.~418425860. This work was performed using the HPC resources provided by CALMIP (Grant 2022/2025-[P21008]).

	\begin{suppinfo}
The Supporting Information is available free of charge at \url{https://pubs.acs.org/}. Methodology and computational details on first-principles calculations and atomistic spin simulations, the interfacial DMI mechanism, the choice of single-\textbf{q} reference states for the estimation of HOI constants, skyrmion size estimation, topological charge density, ferric transitions at finite magnetic fields, and energy density maps with and without HOI.
	\end{suppinfo}
	
	\bibliography{References}

\providecommand{\latin}[1]{#1}
\makeatletter
\providecommand{\doi}
  {\begingroup\let\do\@makeother\dospecials
  \catcode`\{=1 \catcode`\}=2 \doi@aux}
\providecommand{\doi@aux}[1]{\endgroup\texttt{#1}}
\makeatother
\providecommand*\mcitethebibliography{\thebibliography}
\csname @ifundefined\endcsname{endmcitethebibliography}
  {\let\endmcitethebibliography\endthebibliography}{}
\begin{mcitethebibliography}{69}
\providecommand*\natexlab[1]{#1}
\providecommand*\mciteSetBstSublistMode[1]{}
\providecommand*\mciteSetBstMaxWidthForm[2]{}
\providecommand*\mciteBstWouldAddEndPuncttrue
  {\def\EndOfBibitem{\unskip.}}
\providecommand*\mciteBstWouldAddEndPunctfalse
  {\let\EndOfBibitem\relax}
\providecommand*\mciteSetBstMidEndSepPunct[3]{}
\providecommand*\mciteSetBstSublistLabelBeginEnd[3]{}
\providecommand*\EndOfBibitem{}
\mciteSetBstSublistMode{f}
\mciteSetBstMaxWidthForm{subitem}{(\alph{mcitesubitemcount})}
\mciteSetBstSublistLabelBeginEnd
  {\mcitemaxwidthsubitemform\space}
  {\relax}
  {\relax}

\bibitem[Fert \latin{et~al.}(2017)Fert, Reyren, and Cros]{fert2017magnetic}
Fert,~A.; Reyren,~N.; Cros,~V. Magnetic skyrmions: advances in physics and
  potential applications. \emph{Nat. Rev. Mater.} \textbf{2017}, \emph{2},
  1--15\relax
\mciteBstWouldAddEndPuncttrue
\mciteSetBstMidEndSepPunct{\mcitedefaultmidpunct}
{\mcitedefaultendpunct}{\mcitedefaultseppunct}\relax
\EndOfBibitem
\bibitem[Everschor-Sitte \latin{et~al.}(2018)Everschor-Sitte, Masell, Reeve,
  and Kläui]{everschor2018perspective}
Everschor-Sitte,~K.; Masell,~J.; Reeve,~R.~M.; Kläui,~M. Perspective: Magnetic
  skyrmions -- Overview of recent progress in an active research field.
  \emph{J. Appl. Phys.} \textbf{2018}, \emph{124}, 240901\relax
\mciteBstWouldAddEndPuncttrue
\mciteSetBstMidEndSepPunct{\mcitedefaultmidpunct}
{\mcitedefaultendpunct}{\mcitedefaultseppunct}\relax
\EndOfBibitem
\bibitem[Luo \latin{et~al.}(2018)Luo, Song, Li, Zhang, Hong, Yang, Zou, Xu, and
  You]{luo2018reconfigurable}
Luo,~S.; Song,~M.; Li,~X.; Zhang,~Y.; Hong,~J.; Yang,~X.; Zou,~X.; Xu,~N.;
  You,~L. Reconfigurable skyrmion logic gates. \emph{Nano Lett.} \textbf{2018},
  \emph{18}, 1180--1184\relax
\mciteBstWouldAddEndPuncttrue
\mciteSetBstMidEndSepPunct{\mcitedefaultmidpunct}
{\mcitedefaultendpunct}{\mcitedefaultseppunct}\relax
\EndOfBibitem
\bibitem[Li \latin{et~al.}(2021)Li, Kang, Zhang, Nie, Zhou, Wang, and
  Zhao]{li2021magnetic}
Li,~S.; Kang,~W.; Zhang,~X.; Nie,~T.; Zhou,~Y.; Wang,~K.~L.; Zhao,~W. Magnetic
  skyrmions for unconventional computing. \emph{Mater. Horiz.} \textbf{2021},
  \emph{8}, 854--868\relax
\mciteBstWouldAddEndPuncttrue
\mciteSetBstMidEndSepPunct{\mcitedefaultmidpunct}
{\mcitedefaultendpunct}{\mcitedefaultseppunct}\relax
\EndOfBibitem
\bibitem[Psaroudaki and Panagopoulos(2021)Psaroudaki, and
  Panagopoulos]{psaroudaki2021skyrmion}
Psaroudaki,~C.; Panagopoulos,~C. Skyrmion qubits: A new class of quantum logic
  elements based on nanoscale magnetization. \emph{Phys. Rev. Lett.}
  \textbf{2021}, \emph{127}, 067201\relax
\mciteBstWouldAddEndPuncttrue
\mciteSetBstMidEndSepPunct{\mcitedefaultmidpunct}
{\mcitedefaultendpunct}{\mcitedefaultseppunct}\relax
\EndOfBibitem
\bibitem[M\"uhlbauer \latin{et~al.}(2009)M\"uhlbauer, Binz, Jonietz,
  Pfleiderer, Rosch, Neubauer, Georgii, and Boni]{mühlbauer2009skyrmion}
M\"uhlbauer,~S.; Binz,~B.; Jonietz,~F.; Pfleiderer,~C.; Rosch,~A.;
  Neubauer,~A.; Georgii,~R.; Boni,~P. Skyrmion lattice in a chiral magnet.
  \emph{Science} \textbf{2009}, \emph{323}, 915--919\relax
\mciteBstWouldAddEndPuncttrue
\mciteSetBstMidEndSepPunct{\mcitedefaultmidpunct}
{\mcitedefaultendpunct}{\mcitedefaultseppunct}\relax
\EndOfBibitem
\bibitem[Yu \latin{et~al.}(2011)Yu, Kanazawa, Onose, Kimoto, Zhang, Ishiwata,
  Matsui, and Tokura]{yu2011near}
Yu,~X.; Kanazawa,~N.; Onose,~Y.; Kimoto,~K.; Zhang,~W.; Ishiwata,~S.;
  Matsui,~Y.; Tokura,~Y. Near room-temperature formation of a skyrmion crystal
  in thin-films of the helimagnet {FeGe}. \emph{Nat. Mater.} \textbf{2011},
  \emph{10}, 106--109\relax
\mciteBstWouldAddEndPuncttrue
\mciteSetBstMidEndSepPunct{\mcitedefaultmidpunct}
{\mcitedefaultendpunct}{\mcitedefaultseppunct}\relax
\EndOfBibitem
\bibitem[Yang \latin{et~al.}(2012)Yang, Li, Lu, Whangbo, Wei, Gong, and
  Xiang]{yang2012strong}
Yang,~J.-H.; Li,~Z.-L.; Lu,~X.; Whangbo,~M.-H.; Wei,~S.-H.; Gong,~X.; Xiang,~H.
  Strong {D}zyaloshinskii-{M}oriya interaction and origin of ferroelectricity
  in {Cu$_2$OSeO$_3$}. \emph{Phys. Rev. Lett.} \textbf{2012}, \emph{109},
  107203\relax
\mciteBstWouldAddEndPuncttrue
\mciteSetBstMidEndSepPunct{\mcitedefaultmidpunct}
{\mcitedefaultendpunct}{\mcitedefaultseppunct}\relax
\EndOfBibitem
\bibitem[Heinze \latin{et~al.}(2011)Heinze, Von~Bergmann, Menzel, Brede,
  Kubetzka, Wiesendanger, Bihlmayer, and Bl{\"u}gel]{heinze2011spontaneous}
Heinze,~S.; Von~Bergmann,~K.; Menzel,~M.; Brede,~J.; Kubetzka,~A.;
  Wiesendanger,~R.; Bihlmayer,~G.; Bl{\"u}gel,~S. Spontaneous atomic-scale
  magnetic skyrmion lattice in two dimensions. \emph{Nat. Phys.} \textbf{2011},
  \emph{7}, 713--718\relax
\mciteBstWouldAddEndPuncttrue
\mciteSetBstMidEndSepPunct{\mcitedefaultmidpunct}
{\mcitedefaultendpunct}{\mcitedefaultseppunct}\relax
\EndOfBibitem
\bibitem[Romming \latin{et~al.}(2013)Romming, Hanneken, Menzel, Bickel, Wolter,
  von Bergmann, Kubetzka, and Wiesendanger]{romming2013writing}
Romming,~N.; Hanneken,~C.; Menzel,~M.; Bickel,~J.~E.; Wolter,~B.; von
  Bergmann,~K.; Kubetzka,~A.; Wiesendanger,~R. Writing and deleting single
  magnetic skyrmions. \emph{Science} \textbf{2013}, \emph{341}, 636--639\relax
\mciteBstWouldAddEndPuncttrue
\mciteSetBstMidEndSepPunct{\mcitedefaultmidpunct}
{\mcitedefaultendpunct}{\mcitedefaultseppunct}\relax
\EndOfBibitem
\bibitem[Dup{\'e} \latin{et~al.}(2014)Dup{\'e}, Hoffmann, Paillard, and
  Heinze]{dupe2014tailoring}
Dup{\'e},~B.; Hoffmann,~M.; Paillard,~C.; Heinze,~S. Tailoring magnetic
  skyrmions in ultra-thin transition metal films. \emph{Nat. Commun.}
  \textbf{2014}, \emph{5}, 4030\relax
\mciteBstWouldAddEndPuncttrue
\mciteSetBstMidEndSepPunct{\mcitedefaultmidpunct}
{\mcitedefaultendpunct}{\mcitedefaultseppunct}\relax
\EndOfBibitem
\bibitem[Meyer \latin{et~al.}(2019)Meyer, Perini, von Malottki, Kubetzka,
  Wiesendanger, von Bergmann, and Heinze]{meyer2019isolated}
Meyer,~S.; Perini,~M.; von Malottki,~S.; Kubetzka,~A.; Wiesendanger,~R.; von
  Bergmann,~K.; Heinze,~S. Isolated zero field sub-10 nm skyrmions in ultrathin
  {Co} films. \emph{Nat. Commun.} \textbf{2019}, \emph{10}, 3823\relax
\mciteBstWouldAddEndPuncttrue
\mciteSetBstMidEndSepPunct{\mcitedefaultmidpunct}
{\mcitedefaultendpunct}{\mcitedefaultseppunct}\relax
\EndOfBibitem
\bibitem[Moreau-Luchaire \latin{et~al.}(2016)Moreau-Luchaire, Moutafis, Reyren,
  Sampaio, Vaz, Van~Horne, Bouzehouane, Garcia, Deranlot, Warnicke,
  \latin{et~al.} others]{moreau2016additive}
Moreau-Luchaire,~C.; Moutafis,~C.; Reyren,~N.; Sampaio,~J.; Vaz,~C.;
  Van~Horne,~N.; Bouzehouane,~K.; Garcia,~K.; Deranlot,~C.; Warnicke,~P.;
  others Additive interfacial chiral interaction in multilayers for
  stabilization of small individual skyrmions at room temperature. \emph{Nat.
  Nanotechnol.} \textbf{2016}, \emph{11}, 444--448\relax
\mciteBstWouldAddEndPuncttrue
\mciteSetBstMidEndSepPunct{\mcitedefaultmidpunct}
{\mcitedefaultendpunct}{\mcitedefaultseppunct}\relax
\EndOfBibitem
\bibitem[Soumyanarayanan \latin{et~al.}(2017)Soumyanarayanan, Raju,
  Gonzalez~Oyarce, Tan, Im, Petrovi{\'c}, Ho, Khoo, Tran, Gan, \latin{et~al.}
  others]{soumyanarayanan2017tunable}
Soumyanarayanan,~A.; Raju,~M.; Gonzalez~Oyarce,~A.; Tan,~A.~K.; Im,~M.-Y.;
  Petrovi{\'c},~A.~P.; Ho,~P.; Khoo,~K.; Tran,~M.; Gan,~C.; others Tunable
  room-temperature magnetic skyrmions in {Ir/Fe/Co/Pt} multilayers. \emph{Nat.
  Mater.} \textbf{2017}, \emph{16}, 898--904\relax
\mciteBstWouldAddEndPuncttrue
\mciteSetBstMidEndSepPunct{\mcitedefaultmidpunct}
{\mcitedefaultendpunct}{\mcitedefaultseppunct}\relax
\EndOfBibitem
\bibitem[Raju \latin{et~al.}(2019)Raju, Yagil, Soumyanarayanan, Tan, Almoalem,
  Ma, Auslaender, and Panagopoulos]{raju2019evolution}
Raju,~M.; Yagil,~A.; Soumyanarayanan,~A.; Tan,~A.~K.; Almoalem,~A.; Ma,~F.;
  Auslaender,~O.; Panagopoulos,~C. The evolution of skyrmions in {Ir/Fe/Co/Pt}
  multilayers and their topological {Hall} signature. \emph{Nat. Commun.}
  \textbf{2019}, \emph{10}, 696\relax
\mciteBstWouldAddEndPuncttrue
\mciteSetBstMidEndSepPunct{\mcitedefaultmidpunct}
{\mcitedefaultendpunct}{\mcitedefaultseppunct}\relax
\EndOfBibitem
\bibitem[Han \latin{et~al.}(2019)Han, Garlow, Liu, Zhang, Li, DiMarzio, Knight,
  Petrovic, Jariwala, and Zhu]{han2019topological}
Han,~M.-G.; Garlow,~J.~A.; Liu,~Y.; Zhang,~H.; Li,~J.; DiMarzio,~D.;
  Knight,~M.~W.; Petrovic,~C.; Jariwala,~D.; Zhu,~Y. Topological magnetic-spin
  textures in two-dimensional van der {Waals} {Cr$_2$Ge$_2$Te$_6$}. \emph{Nano
  Lett.} \textbf{2019}, \emph{19}, 7859--7865\relax
\mciteBstWouldAddEndPuncttrue
\mciteSetBstMidEndSepPunct{\mcitedefaultmidpunct}
{\mcitedefaultendpunct}{\mcitedefaultseppunct}\relax
\EndOfBibitem
\bibitem[Ding \latin{et~al.}(2019)Ding, Li, Xu, Li, Hou, Liu, Xi, Xu, Yao, and
  Wang]{ding2019observation}
Ding,~B.; Li,~Z.; Xu,~G.; Li,~H.; Hou,~Z.; Liu,~E.; Xi,~X.; Xu,~F.; Yao,~Y.;
  Wang,~W. Observation of magnetic skyrmion bubbles in a van der {Waals}
  ferromagnet {Fe$_3$GeTe$_2$}. \emph{Nano Lett.} \textbf{2019}, \emph{20},
  868--873\relax
\mciteBstWouldAddEndPuncttrue
\mciteSetBstMidEndSepPunct{\mcitedefaultmidpunct}
{\mcitedefaultendpunct}{\mcitedefaultseppunct}\relax
\EndOfBibitem
\bibitem[Schmitt \latin{et~al.}(2022)Schmitt, \latin{et~al.}
  others]{schmitt2022skyrmionic}
Schmitt,~M.; others Skyrmionic spin structures in layered Fe$_5$GeTe$_2$ up to
  room temperature. \emph{Commun. Phys.} \textbf{2022}, \emph{5}, 254\relax
\mciteBstWouldAddEndPuncttrue
\mciteSetBstMidEndSepPunct{\mcitedefaultmidpunct}
{\mcitedefaultendpunct}{\mcitedefaultseppunct}\relax
\EndOfBibitem
\bibitem[Zhang \latin{et~al.}(2024)Zhang, Lu, Tabrizian, Feng, and
  Wu]{zhang20242d}
Zhang,~B.; Lu,~P.; Tabrizian,~R.; Feng,~P. X.-L.; Wu,~Y. 2D magnetic
  heterostructures: Spintronics and quantum future. \emph{npj Spintronics}
  \textbf{2024}, \emph{2}, 6\relax
\mciteBstWouldAddEndPuncttrue
\mciteSetBstMidEndSepPunct{\mcitedefaultmidpunct}
{\mcitedefaultendpunct}{\mcitedefaultseppunct}\relax
\EndOfBibitem
\bibitem[Li \latin{et~al.}(2022)Li, Haldar, and Heinze]{li2022strain}
Li,~D.; Haldar,~S.; Heinze,~S. Strain-driven zero-field near-10 nm skyrmions in
  two-dimensional van der {Waals} heterostructures. \emph{Nano Lett.}
  \textbf{2022}, \emph{22}, 7706--7713\relax
\mciteBstWouldAddEndPuncttrue
\mciteSetBstMidEndSepPunct{\mcitedefaultmidpunct}
{\mcitedefaultendpunct}{\mcitedefaultseppunct}\relax
\EndOfBibitem
\bibitem[Wu \latin{et~al.}(2020)Wu, Zhang, Zhang, Wang, Zhu, Hu, Yin, Wong,
  Fang, Wan, \latin{et~al.} others]{wu2020neel}
Wu,~Y.; Zhang,~S.; Zhang,~J.; Wang,~W.; Zhu,~Y.~L.; Hu,~J.; Yin,~G.; Wong,~K.;
  Fang,~C.; Wan,~C.; others N{\'e}el-type skyrmion in {WTe$_2$/Fe$_3$GeTe$_2$}
  van der {Waals} heterostructure. \emph{Nat. Commun.} \textbf{2020},
  \emph{11}, 3860\relax
\mciteBstWouldAddEndPuncttrue
\mciteSetBstMidEndSepPunct{\mcitedefaultmidpunct}
{\mcitedefaultendpunct}{\mcitedefaultseppunct}\relax
\EndOfBibitem
\bibitem[Li \latin{et~al.}(2023)Li, Haldar, Drevelow, and
  Heinze]{Dongzhe_prb2023}
Li,~D.; Haldar,~S.; Drevelow,~T.; Heinze,~S. Tuning the magnetic interactions
  in van der {W}aals {${\mathrm{Fe}}_{3}{\mathrm{GeTe}}_{2}$} heterostructures:
  A comparative study of \textit{ab initio} methods. \emph{Phys. Rev. B}
  \textbf{2023}, \emph{107}, 104428\relax
\mciteBstWouldAddEndPuncttrue
\mciteSetBstMidEndSepPunct{\mcitedefaultmidpunct}
{\mcitedefaultendpunct}{\mcitedefaultseppunct}\relax
\EndOfBibitem
\bibitem[Sun \latin{et~al.}(2020)Sun, Wang, Li, Zhang, Chen, Wang, and
  Cheng]{sun2020controlling}
Sun,~W.; Wang,~W.; Li,~H.; Zhang,~G.; Chen,~D.; Wang,~J.; Cheng,~Z. Controlling
  bimerons as skyrmion analogues by ferroelectric polarization in {2D} van der
  {Waals} multiferroic heterostructures. \emph{Nat. Commun.} \textbf{2020},
  \emph{11}, 5930\relax
\mciteBstWouldAddEndPuncttrue
\mciteSetBstMidEndSepPunct{\mcitedefaultmidpunct}
{\mcitedefaultendpunct}{\mcitedefaultseppunct}\relax
\EndOfBibitem
\bibitem[Huang \latin{et~al.}(2022)Huang, Shao, and Tsymbal]{Huang2022}
Huang,~K.; Shao,~D.-F.; Tsymbal,~E.~Y. Ferroelectric control of magnetic
  skyrmions in two-dimensional van der {Waals} heterostructures. \emph{Nano
  Lett.} \textbf{2022}, \emph{22}, 3349--3355\relax
\mciteBstWouldAddEndPuncttrue
\mciteSetBstMidEndSepPunct{\mcitedefaultmidpunct}
{\mcitedefaultendpunct}{\mcitedefaultseppunct}\relax
\EndOfBibitem
\bibitem[Tong \latin{et~al.}(2018)Tong, Liu, Xiao, and Yao]{Tong2018}
Tong,~Q.; Liu,~F.; Xiao,~J.; Yao,~W. Skyrmions in the {Moiré} of van der
  {Waals} {2D} magnets. \emph{Nano Lett.} \textbf{2018}, \emph{18},
  7194--7199\relax
\mciteBstWouldAddEndPuncttrue
\mciteSetBstMidEndSepPunct{\mcitedefaultmidpunct}
{\mcitedefaultendpunct}{\mcitedefaultseppunct}\relax
\EndOfBibitem
\bibitem[Khela \latin{et~al.}(2023)Khela, Dabrowski, Khan, Keatley,
  Verzhbitskiy, Eda, Hicken, Kurebayashi, and Santos]{khela2023laser}
Khela,~M.; Dabrowski,~M.; Khan,~S.; Keatley,~P.~S.; Verzhbitskiy,~I.; Eda,~G.;
  Hicken,~R.~J.; Kurebayashi,~H.; Santos,~E.~J. Laser-induced topological spin
  switching in a {2D} van der {W}aals magnet. \emph{Nat. Commun.}
  \textbf{2023}, \emph{14}, 1378\relax
\mciteBstWouldAddEndPuncttrue
\mciteSetBstMidEndSepPunct{\mcitedefaultmidpunct}
{\mcitedefaultendpunct}{\mcitedefaultseppunct}\relax
\EndOfBibitem
\bibitem[Li \latin{et~al.}(2024)Li, Haldar, and Heinze]{li2024proposal}
Li,~D.; Haldar,~S.; Heinze,~S. Proposal for all-electrical skyrmion detection
  in van der {Waals} tunnel junctions. \emph{Nano Lett.} \textbf{2024},
  \emph{24}, 2496--2502\relax
\mciteBstWouldAddEndPuncttrue
\mciteSetBstMidEndSepPunct{\mcitedefaultmidpunct}
{\mcitedefaultendpunct}{\mcitedefaultseppunct}\relax
\EndOfBibitem
\bibitem[Polyakov and Belavin()Polyakov, and Belavin]{polyakov22metastable}
Polyakov,~A.; Belavin,~A. Metastable states of two-dimensional isotropic
  ferromagnets 1975. \emph{Sov. Phys. JETP Lett} \emph{22}, 245\relax
\mciteBstWouldAddEndPuncttrue
\mciteSetBstMidEndSepPunct{\mcitedefaultmidpunct}
{\mcitedefaultendpunct}{\mcitedefaultseppunct}\relax
\EndOfBibitem
\bibitem[Bernand-Mantel \latin{et~al.}(2022)Bernand-Mantel, Muratov, and
  Slastikov]{bernand2022micromagnetic}
Bernand-Mantel,~A.; Muratov,~C.~B.; Slastikov,~V.~V. A micromagnetic theory of
  skyrmion lifetime in ultrathin ferromagnetic films. \emph{Proc. Natl. Acad.
  Sci. U.S.A.} \textbf{2022}, \emph{119}, e2122237119\relax
\mciteBstWouldAddEndPuncttrue
\mciteSetBstMidEndSepPunct{\mcitedefaultmidpunct}
{\mcitedefaultendpunct}{\mcitedefaultseppunct}\relax
\EndOfBibitem
\bibitem[B{\"u}ttner \latin{et~al.}(2018)B{\"u}ttner, Lemesh, and
  Beach]{buttner2018theory}
B{\"u}ttner,~F.; Lemesh,~I.; Beach,~G.~S. Theory of isolated magnetic
  skyrmions: From fundamentals to room temperature applications. \emph{Sci.
  Rep.} \textbf{2018}, \emph{8}, 4464\relax
\mciteBstWouldAddEndPuncttrue
\mciteSetBstMidEndSepPunct{\mcitedefaultmidpunct}
{\mcitedefaultendpunct}{\mcitedefaultseppunct}\relax
\EndOfBibitem
\bibitem[Leonov \latin{et~al.}(2016)Leonov, Monchesky, Romming, Kubetzka,
  Bogdanov, and Wiesendanger]{leonov2016properties}
Leonov,~A.; Monchesky,~T.; Romming,~N.; Kubetzka,~A.; Bogdanov,~A.;
  Wiesendanger,~R. The properties of isolated chiral skyrmions in thin magnetic
  films. \emph{New J. Phys.} \textbf{2016}, \emph{18}, 065003\relax
\mciteBstWouldAddEndPuncttrue
\mciteSetBstMidEndSepPunct{\mcitedefaultmidpunct}
{\mcitedefaultendpunct}{\mcitedefaultseppunct}\relax
\EndOfBibitem
\bibitem[Muckel \latin{et~al.}(2021)Muckel, von Malottki, Holl, Pestka,
  Pratzer, Bessarab, Heinze, and Morgenstern]{muckel2021experimental}
Muckel,~F.; von Malottki,~S.; Holl,~C.; Pestka,~B.; Pratzer,~M.;
  Bessarab,~P.~F.; Heinze,~S.; Morgenstern,~M. Experimental identification of
  two distinct skyrmion collapse mechanisms. \emph{Nat. Phys.} \textbf{2021},
  \emph{17}, 395--402\relax
\mciteBstWouldAddEndPuncttrue
\mciteSetBstMidEndSepPunct{\mcitedefaultmidpunct}
{\mcitedefaultendpunct}{\mcitedefaultseppunct}\relax
\EndOfBibitem
\bibitem[Li \latin{et~al.}(2024)Li, Goerzen, Haldar, Drevelow, Schrautzer, and
  Heinze]{li2024stability}
Li,~D.; Goerzen,~M.~A.; Haldar,~S.; Drevelow,~T.; Schrautzer,~H.; Heinze,~S.
  Stability and localization of nanoscale skyrmions and bimerons in an
  all-magnetic van der {Waals} heterostructure. \emph{arXiv preprint
  arXiv:2408.15974} \textbf{2024}, \relax
\mciteBstWouldAddEndPunctfalse
\mciteSetBstMidEndSepPunct{\mcitedefaultmidpunct}
{}{\mcitedefaultseppunct}\relax
\EndOfBibitem
\bibitem[Liang \latin{et~al.}(2020)Liang, Wang, Du, Hallal, Garcia, Chshiev,
  Fert, and Yang]{liang2020very}
Liang,~J.; Wang,~W.; Du,~H.; Hallal,~A.; Garcia,~K.; Chshiev,~M.; Fert,~A.;
  Yang,~H. Very large {D}zyaloshinskii-{M}oriya interaction in two-dimensional
  {J}anus manganese dichalcogenides and its application to realize skyrmion
  states. \emph{Phys. Rev. B} \textbf{2020}, \emph{101}, 184401\relax
\mciteBstWouldAddEndPuncttrue
\mciteSetBstMidEndSepPunct{\mcitedefaultmidpunct}
{\mcitedefaultendpunct}{\mcitedefaultseppunct}\relax
\EndOfBibitem
\bibitem[Li \latin{et~al.}(2023)Li, Yu, Liang, Ga, and Yang]{li2023topological}
Li,~P.; Yu,~D.; Liang,~J.; Ga,~Y.; Yang,~H. Topological spin textures in 1
  \textit{T}-phase {J}anus magnets: Interplay between {Dzyaloshinskii-Moriya}
  interaction, magnetic frustration, and isotropic higher-order interactions.
  \emph{Phys. Rev. B} \textbf{2023}, \emph{107}, 054408\relax
\mciteBstWouldAddEndPuncttrue
\mciteSetBstMidEndSepPunct{\mcitedefaultmidpunct}
{\mcitedefaultendpunct}{\mcitedefaultseppunct}\relax
\EndOfBibitem
\bibitem[Li \latin{et~al.}(2024)Li, Haldar, Kollwitz, Schrautzer, Goerzen, and
  Heinze]{Dongzhe_PRB2024}
Li,~D.; Haldar,~S.; Kollwitz,~L.; Schrautzer,~H.; Goerzen,~M.~A.; Heinze,~S.
  Prediction of stable nanoscale skyrmions in monolayer {Fe$_5$GeTe$_2$}.
  \emph{Phys. Rev. B} \textbf{2024}, \emph{109}, L220404\relax
\mciteBstWouldAddEndPuncttrue
\mciteSetBstMidEndSepPunct{\mcitedefaultmidpunct}
{\mcitedefaultendpunct}{\mcitedefaultseppunct}\relax
\EndOfBibitem
\bibitem[Takahashi(1977)]{takahashi1977half}
Takahashi,~M. Half-filled {H}ubbard model at low temperature. \emph{J. Phys. C:
  Solid State Phys.} \textbf{1977}, \emph{10}, 1289\relax
\mciteBstWouldAddEndPuncttrue
\mciteSetBstMidEndSepPunct{\mcitedefaultmidpunct}
{\mcitedefaultendpunct}{\mcitedefaultseppunct}\relax
\EndOfBibitem
\bibitem[MacDonald \latin{et~al.}(1988)MacDonald, Girvin, and
  Yoshioka]{macdonald1988t}
MacDonald,~A.~H.; Girvin,~S.~M.; Yoshioka,~D. $\frac{t}{U}$ expansion for the
  Hubbard model. \emph{Phys. Rev. B} \textbf{1988}, \emph{37}, 9753--9756\relax
\mciteBstWouldAddEndPuncttrue
\mciteSetBstMidEndSepPunct{\mcitedefaultmidpunct}
{\mcitedefaultendpunct}{\mcitedefaultseppunct}\relax
\EndOfBibitem
\bibitem[Hoffmann and Bl{\"u}gel(2020)Hoffmann, and
  Bl{\"u}gel]{hoffmann2020systematic}
Hoffmann,~M.; Bl{\"u}gel,~S. Systematic derivation of realistic spin models for
  beyond-{H}eisenberg solids. \emph{Phys. Rev. B} \textbf{2020}, \emph{101},
  024418\relax
\mciteBstWouldAddEndPuncttrue
\mciteSetBstMidEndSepPunct{\mcitedefaultmidpunct}
{\mcitedefaultendpunct}{\mcitedefaultseppunct}\relax
\EndOfBibitem
\bibitem[Kurz \latin{et~al.}(2001)Kurz, Bihlmayer, Hirai, and
  Bl{\"u}gel]{kurz2001three}
Kurz,~P.; Bihlmayer,~G.; Hirai,~K.; Bl{\"u}gel,~S. Three-dimensional spin
  structure on a two-dimensional lattice: {Mn/Cu} (111). \emph{Phys. Rev.
  Lett.} \textbf{2001}, \emph{86}, 1106\relax
\mciteBstWouldAddEndPuncttrue
\mciteSetBstMidEndSepPunct{\mcitedefaultmidpunct}
{\mcitedefaultendpunct}{\mcitedefaultseppunct}\relax
\EndOfBibitem
\bibitem[Kr{\"o}nlein \latin{et~al.}(2018)Kr{\"o}nlein, Schmitt, Hoffmann,
  Kemmer, Seubert, Vogt, K{\"u}spert, B{\"o}hme, Alonazi, K{\"u}gel,
  \latin{et~al.} others]{kronlein2018magnetic}
Kr{\"o}nlein,~A.; Schmitt,~M.; Hoffmann,~M.; Kemmer,~J.; Seubert,~N.; Vogt,~M.;
  K{\"u}spert,~J.; B{\"o}hme,~M.; Alonazi,~B.; K{\"u}gel,~J.; others Magnetic
  ground state stabilized by three-site interactions: {Fe/Rh} (111).
  \emph{Phys. Rev. Lett.} \textbf{2018}, \emph{120}, 207202\relax
\mciteBstWouldAddEndPuncttrue
\mciteSetBstMidEndSepPunct{\mcitedefaultmidpunct}
{\mcitedefaultendpunct}{\mcitedefaultseppunct}\relax
\EndOfBibitem
\bibitem[Spethmann \latin{et~al.}(2020)Spethmann, Meyer, von Bergmann,
  Wiesendanger, Heinze, and Kubetzka]{spethmann2020discovery}
Spethmann,~J.; Meyer,~S.; von Bergmann,~K.; Wiesendanger,~R.; Heinze,~S.;
  Kubetzka,~A. Discovery of magnetic single-and triple-q states in {Mn/Re}
  (0001). \emph{Phys. Rev. Lett.} \textbf{2020}, \emph{124}, 227203\relax
\mciteBstWouldAddEndPuncttrue
\mciteSetBstMidEndSepPunct{\mcitedefaultmidpunct}
{\mcitedefaultendpunct}{\mcitedefaultseppunct}\relax
\EndOfBibitem
\bibitem[Kartsev \latin{et~al.}(2020)Kartsev, Augustin, Evans, Novoselov, and
  Santos]{kartsev2020biquadratic}
Kartsev,~A.; Augustin,~M.; Evans,~R.~F.; Novoselov,~K.~S.; Santos,~E.~J.
  Biquadratic exchange interactions in two-dimensional magnets. \emph{npj
  Comput. Mater.} \textbf{2020}, \emph{6}, 150\relax
\mciteBstWouldAddEndPuncttrue
\mciteSetBstMidEndSepPunct{\mcitedefaultmidpunct}
{\mcitedefaultendpunct}{\mcitedefaultseppunct}\relax
\EndOfBibitem
\bibitem[Paul \latin{et~al.}(2020)Paul, Haldar, von Malottki, and
  Heinze]{paul2020role}
Paul,~S.; Haldar,~S.; von Malottki,~S.; Heinze,~S. Role of higher-order
  exchange interactions for skyrmion stability. \emph{Nat. Commun.}
  \textbf{2020}, \emph{11}, 4756\relax
\mciteBstWouldAddEndPuncttrue
\mciteSetBstMidEndSepPunct{\mcitedefaultmidpunct}
{\mcitedefaultendpunct}{\mcitedefaultseppunct}\relax
\EndOfBibitem
\bibitem[Haldar \latin{et~al.}(2021)Haldar, Meyer, Kubetzka, and
  Heinze]{haldar2021}
Haldar,~S.; Meyer,~S.; Kubetzka,~A.; Heinze,~S. {Distorted 3Q state driven by
  topological-chiral magnetic interactions}. \emph{Phys. Rev. B} \textbf{2021},
  \emph{104}, L180404\relax
\mciteBstWouldAddEndPuncttrue
\mciteSetBstMidEndSepPunct{\mcitedefaultmidpunct}
{\mcitedefaultendpunct}{\mcitedefaultseppunct}\relax
\EndOfBibitem
\bibitem[Gutzeit \latin{et~al.}(2022)Gutzeit, Kubetzka, Haldar, Pralow,
  Goerzen, Wiesendanger, Heinze, and von Bergmann]{gutzeit2022nano}
Gutzeit,~M.; Kubetzka,~A.; Haldar,~S.; Pralow,~H.; Goerzen,~M.~A.;
  Wiesendanger,~R.; Heinze,~S.; von Bergmann,~K. Nano-scale collinear multi-{Q}
  states driven by higher-order interactions. \emph{Nat. Commun.}
  \textbf{2022}, \emph{13}, 5764\relax
\mciteBstWouldAddEndPuncttrue
\mciteSetBstMidEndSepPunct{\mcitedefaultmidpunct}
{\mcitedefaultendpunct}{\mcitedefaultseppunct}\relax
\EndOfBibitem
\bibitem[Xu \latin{et~al.}(2022)Xu, Li, Chen, Zhang, Xiang, and
  Bellaiche]{xu2022assembling}
Xu,~C.; Li,~X.; Chen,~P.; Zhang,~Y.; Xiang,~H.; Bellaiche,~L. Assembling
  diverse skyrmionic phases in Fe$_3$GeTe$_2$ monolayers. \emph{Adv. Mater.}
  \textbf{2022}, \emph{34}, 2107779\relax
\mciteBstWouldAddEndPuncttrue
\mciteSetBstMidEndSepPunct{\mcitedefaultmidpunct}
{\mcitedefaultendpunct}{\mcitedefaultseppunct}\relax
\EndOfBibitem
\bibitem[Pan \latin{et~al.}(2024)Pan, Xu, Li, Xu, Liu, Gu, and
  Duan]{pan2024chiral}
Pan,~W.; Xu,~C.; Li,~X.; Xu,~Z.; Liu,~B.; Gu,~B.-L.; Duan,~W. Chiral magnetism
  in lithium-decorated monolayer {CrTe$_2$}: Interplay between
  {Dzyaloshinskii-Moriya} interaction and higher-order interactions.
  \emph{Phys. Rev. B} \textbf{2024}, \emph{109}, 214405\relax
\mciteBstWouldAddEndPuncttrue
\mciteSetBstMidEndSepPunct{\mcitedefaultmidpunct}
{\mcitedefaultendpunct}{\mcitedefaultseppunct}\relax
\EndOfBibitem
\bibitem[Yuan \latin{et~al.}(2020)Yuan, Yang, Cai, Wu, Chen, Yan, and
  Shen]{yuan2020intrinsic}
Yuan,~J.; Yang,~Y.; Cai,~Y.; Wu,~Y.; Chen,~Y.; Yan,~X.; Shen,~L. Intrinsic
  skyrmions in monolayer {J}anus magnets. \emph{Phys. Rev. B} \textbf{2020},
  \emph{101}, 094420\relax
\mciteBstWouldAddEndPuncttrue
\mciteSetBstMidEndSepPunct{\mcitedefaultmidpunct}
{\mcitedefaultendpunct}{\mcitedefaultseppunct}\relax
\EndOfBibitem
\bibitem[Xu \latin{et~al.}(2020)Xu, Feng, Prokhorenko, Nahas, Xiang, and
  Bellaiche]{changsong2020}
Xu,~C.; Feng,~J.; Prokhorenko,~S.; Nahas,~Y.; Xiang,~H.; Bellaiche,~L.
  Topological spin texture in {Janus} monolayers of the chromium trihalides
  {Cr(I, ${X)}_{3}$}. \emph{Phys. Rev. B} \textbf{2020}, \emph{101},
  060404\relax
\mciteBstWouldAddEndPuncttrue
\mciteSetBstMidEndSepPunct{\mcitedefaultmidpunct}
{\mcitedefaultendpunct}{\mcitedefaultseppunct}\relax
\EndOfBibitem
\bibitem[Shen \latin{et~al.}(2022)Shen, Song, Xue, Wu, Wang, and
  Zhong]{changsheng2022}
Shen,~Z.; Song,~C.; Xue,~Y.; Wu,~Z.; Wang,~J.; Zhong,~Z. Strain-tunable
  {Dzyaloshinskii-Moriya} interaction and skyrmions in two-dimensional {Janus
  ${\mathrm{Cr}}_{2}{X}_{3}{Y}_{3}$ ($X, Y$ = Cl, Br, I, $X\ensuremath{\ne}Y$)}
  trihalide monolayers. \emph{Phys. Rev. B} \textbf{2022}, \emph{106},
  094403\relax
\mciteBstWouldAddEndPuncttrue
\mciteSetBstMidEndSepPunct{\mcitedefaultmidpunct}
{\mcitedefaultendpunct}{\mcitedefaultseppunct}\relax
\EndOfBibitem
\bibitem[Zhou \latin{et~al.}(2024)Zhou, Shaikh, Sun, Chen, Chen, Li, Tong,
  Sanyal, and Wang]{zhou2024emergence}
Zhou,~F.; Shaikh,~M.; Sun,~W.; Chen,~F.; Chen,~X.; Li,~S.; Tong,~H.;
  Sanyal,~B.; Wang,~D. Emergence of polar skyrmions in 2D {Janus CrIn$X_3$
  ($X$= Se, Te)} magnets. \emph{npj 2D Mater. Appl.} \textbf{2024}, \emph{8},
  51\relax
\mciteBstWouldAddEndPuncttrue
\mciteSetBstMidEndSepPunct{\mcitedefaultmidpunct}
{\mcitedefaultendpunct}{\mcitedefaultseppunct}\relax
\EndOfBibitem
\bibitem[Xu \latin{et~al.}(2025)Xu, Shao, Huang, Zhu, Hu, Hu, Li, Hao, Hou,
  Zhang, \latin{et~al.} others]{xu2025unusual}
Xu,~Z.; Shao,~Y.; Huang,~C.; Zhu,~C.; Hu,~G.; Hu,~S.; Li,~Z.-L.; Hao,~X.;
  Hou,~Y.; Zhang,~T.; others Unusual charge density wave introduced by the
  Janus structure in monolayer vanadium dichalcogenides. \emph{Sci. Adv.}
  \textbf{2025}, \emph{11}, eadq4406\relax
\mciteBstWouldAddEndPuncttrue
\mciteSetBstMidEndSepPunct{\mcitedefaultmidpunct}
{\mcitedefaultendpunct}{\mcitedefaultseppunct}\relax
\EndOfBibitem
\bibitem[Nie \latin{et~al.}(2024)Nie, \latin{et~al.} others]{nie2024regulated}
Nie,~J.-H.; others {Regulated magnetic anisotropy and charge density wave in
  uniformly fabricated Janus CrTeSe monolayer}. \emph{arXiv:} \textbf{2024},
  \emph{2407.16569}\relax
\mciteBstWouldAddEndPuncttrue
\mciteSetBstMidEndSepPunct{\mcitedefaultmidpunct}
{\mcitedefaultendpunct}{\mcitedefaultseppunct}\relax
\EndOfBibitem
\bibitem[von Malottki \latin{et~al.}(2017)von Malottki, Dup{\'e}, Bessarab,
  Delin, and Heinze]{von2017enhanced}
von Malottki,~S.; Dup{\'e},~B.; Bessarab,~P.~F.; Delin,~A.; Heinze,~S. Enhanced
  skyrmion stability due to exchange frustration. \emph{Sci. Rep.}
  \textbf{2017}, \emph{7}, 12299\relax
\mciteBstWouldAddEndPuncttrue
\mciteSetBstMidEndSepPunct{\mcitedefaultmidpunct}
{\mcitedefaultendpunct}{\mcitedefaultseppunct}\relax
\EndOfBibitem
\bibitem[Goerzen \latin{et~al.}(2023)Goerzen, von Malottki, Meyer, Bessarab,
  and Heinze]{goerzen2023lifetime}
Goerzen,~M.~A.; von Malottki,~S.; Meyer,~S.; Bessarab,~P.~F.; Heinze,~S.
  Lifetime of coexisting sub-10 nm zero-field skyrmions and antiskyrmions.
  \emph{npj Quantum Mater.} \textbf{2023}, \emph{8}, 54\relax
\mciteBstWouldAddEndPuncttrue
\mciteSetBstMidEndSepPunct{\mcitedefaultmidpunct}
{\mcitedefaultendpunct}{\mcitedefaultseppunct}\relax
\EndOfBibitem
\bibitem[R\'ozsa \latin{et~al.}(2017)R\'ozsa, Palot\'as, De\'ak, Simon, Yanes,
  Udvardi, Szunyogh, and Nowak]{Levente2017}
R\'ozsa,~L.; Palot\'as,~K.; De\'ak,~A.; Simon,~E.; Yanes,~R.; Udvardi,~L.;
  Szunyogh,~L.; Nowak,~U. Formation and stability of metastable skyrmionic spin
  structures with various topologies in an ultrathin film. \emph{Phys. Rev. B}
  \textbf{2017}, \emph{95}, 094423\relax
\mciteBstWouldAddEndPuncttrue
\mciteSetBstMidEndSepPunct{\mcitedefaultmidpunct}
{\mcitedefaultendpunct}{\mcitedefaultseppunct}\relax
\EndOfBibitem
\bibitem[fle()]{fleurv26}
www.flapw.de (accessed January 1, 2024)\relax
\mciteBstWouldAddEndPuncttrue
\mciteSetBstMidEndSepPunct{\mcitedefaultmidpunct}
{\mcitedefaultendpunct}{\mcitedefaultseppunct}\relax
\EndOfBibitem
\bibitem[Kurz \latin{et~al.}(2004)Kurz, F{\"o}rster, Nordstr{\"o}m, Bihlmayer,
  and Bl{\"u}gel]{kurz2004ab}
Kurz,~P.; F{\"o}rster,~F.; Nordstr{\"o}m,~L.; Bihlmayer,~G.; Bl{\"u}gel,~S.
  \textit{Ab-initio} treatment of noncollinear magnets with the full-potential
  linearized augmented plane wave method. \emph{Phys. Rev. B} \textbf{2004},
  \emph{69}, 024415\relax
\mciteBstWouldAddEndPuncttrue
\mciteSetBstMidEndSepPunct{\mcitedefaultmidpunct}
{\mcitedefaultendpunct}{\mcitedefaultseppunct}\relax
\EndOfBibitem
\bibitem[Heide \latin{et~al.}(2009)Heide, Bihlmayer, and
  Bl{\"u}gel]{heide2009describing}
Heide,~M.; Bihlmayer,~G.; Bl{\"u}gel,~S. Describing {Dzyaloshinskii--Moriya}
  spirals from first principles. \emph{Phys. B: Condens. Matter.}
  \textbf{2009}, \emph{404}, 2678--2683\relax
\mciteBstWouldAddEndPuncttrue
\mciteSetBstMidEndSepPunct{\mcitedefaultmidpunct}
{\mcitedefaultendpunct}{\mcitedefaultseppunct}\relax
\EndOfBibitem
\bibitem[Fert and Levy(1980)Fert, and Levy]{Fert1980}
Fert,~A.; Levy,~P.~M. Role of anisotropic exchange interactions in determining
  the properties of spin-glasses. \emph{Phys. Rev. Lett.} \textbf{1980},
  \emph{44}, 1538--1541\relax
\mciteBstWouldAddEndPuncttrue
\mciteSetBstMidEndSepPunct{\mcitedefaultmidpunct}
{\mcitedefaultendpunct}{\mcitedefaultseppunct}\relax
\EndOfBibitem
\bibitem[Hardrat \latin{et~al.}(2009)Hardrat, Al-Zubi, Ferriani, Bl{\"u}gel,
  Bihlmayer, and Heinze]{hardrat2009complex}
Hardrat,~B.; Al-Zubi,~A.; Ferriani,~P.; Bl{\"u}gel,~S.; Bihlmayer,~G.;
  Heinze,~S. Complex magnetism of iron monolayers on hexagonal transition metal
  surfaces from first principles. \emph{Phys. Rev. B} \textbf{2009}, \emph{79},
  094411\relax
\mciteBstWouldAddEndPuncttrue
\mciteSetBstMidEndSepPunct{\mcitedefaultmidpunct}
{\mcitedefaultendpunct}{\mcitedefaultseppunct}\relax
\EndOfBibitem
\bibitem[Tan \latin{et~al.}(2018)Tan, Lee, Jung, Park, Albarakati, Partridge,
  Field, McCulloch, Wang, and Lee]{tan2018hard}
Tan,~C.; Lee,~J.; Jung,~S.-G.; Park,~T.; Albarakati,~S.; Partridge,~J.;
  Field,~M.~R.; McCulloch,~D.~G.; Wang,~L.; Lee,~C. Hard magnetic properties in
  nanoflake van der {Waals} {Fe$_3$GeTe$_2$}. \emph{Nat. Commun.}
  \textbf{2018}, \emph{9}, 1554\relax
\mciteBstWouldAddEndPuncttrue
\mciteSetBstMidEndSepPunct{\mcitedefaultmidpunct}
{\mcitedefaultendpunct}{\mcitedefaultseppunct}\relax
\EndOfBibitem
\bibitem[Wang \latin{et~al.}(2020)Wang, Chen, and Long]{wang2020modifications}
Wang,~Y.-P.; Chen,~X.-Y.; Long,~M.-Q. Modifications of magnetic anisotropy of
  {Fe$_3$GeTe$_2$} by the electric field effect. \emph{Appl. Phys. Lett.}
  \textbf{2020}, \emph{116}, 092404\relax
\mciteBstWouldAddEndPuncttrue
\mciteSetBstMidEndSepPunct{\mcitedefaultmidpunct}
{\mcitedefaultendpunct}{\mcitedefaultseppunct}\relax
\EndOfBibitem
\bibitem[Park \latin{et~al.}(2019)Park, Kim, Liu, Hwang, Kim, Kim, Kim,
  Petrovic, Hwang, Mo, \latin{et~al.} others]{park2019controlling}
Park,~S.~Y.; Kim,~D.~S.; Liu,~Y.; Hwang,~J.; Kim,~Y.; Kim,~W.; Kim,~J.-Y.;
  Petrovic,~C.; Hwang,~C.; Mo,~S.-K.; others Controlling the magnetic
  anisotropy of the van der {W}aals ferromagnet {Fe$_3$GeTe$_2$} through hole
  doping. \emph{Nano Lett.} \textbf{2019}, \emph{20}, 95--100\relax
\mciteBstWouldAddEndPuncttrue
\mciteSetBstMidEndSepPunct{\mcitedefaultmidpunct}
{\mcitedefaultendpunct}{\mcitedefaultseppunct}\relax
\EndOfBibitem
\bibitem[Bocdanov and Hubert(1994)Bocdanov, and Hubert]{bocdanov1994properties}
Bocdanov,~A.; Hubert,~A. The properties of isolated magnetic vortices.
  \emph{Phys. Status Solidi B} \textbf{1994}, \emph{186}, 527--543\relax
\mciteBstWouldAddEndPuncttrue
\mciteSetBstMidEndSepPunct{\mcitedefaultmidpunct}
{\mcitedefaultendpunct}{\mcitedefaultseppunct}\relax
\EndOfBibitem
\bibitem[Wang \latin{et~al.}(2018)Wang, Yuan, and Wang]{wang2018theory}
Wang,~X.; Yuan,~H.; Wang,~X. A theory on skyrmion size. \emph{Commun. Phys.}
  \textbf{2018}, \emph{1}, 31\relax
\mciteBstWouldAddEndPuncttrue
\mciteSetBstMidEndSepPunct{\mcitedefaultmidpunct}
{\mcitedefaultendpunct}{\mcitedefaultseppunct}\relax
\EndOfBibitem
\bibitem[Bessarab \latin{et~al.}(2015)Bessarab, Uzdin, and
  J{\'o}nsson]{bessarab2015method}
Bessarab,~P.~F.; Uzdin,~V.~M.; J{\'o}nsson,~H. Method for finding mechanism and
  activation energy of magnetic transitions, applied to skyrmion and antivortex
  annihilation. \emph{Comput. Phys. Commun.} \textbf{2015}, \emph{196},
  335--347\relax
\mciteBstWouldAddEndPuncttrue
\mciteSetBstMidEndSepPunct{\mcitedefaultmidpunct}
{\mcitedefaultendpunct}{\mcitedefaultseppunct}\relax
\EndOfBibitem
\end{mcitethebibliography}

\newpage

\begin{figure*}[tbp]
    \centering
    \includegraphics[width=1.0\linewidth]{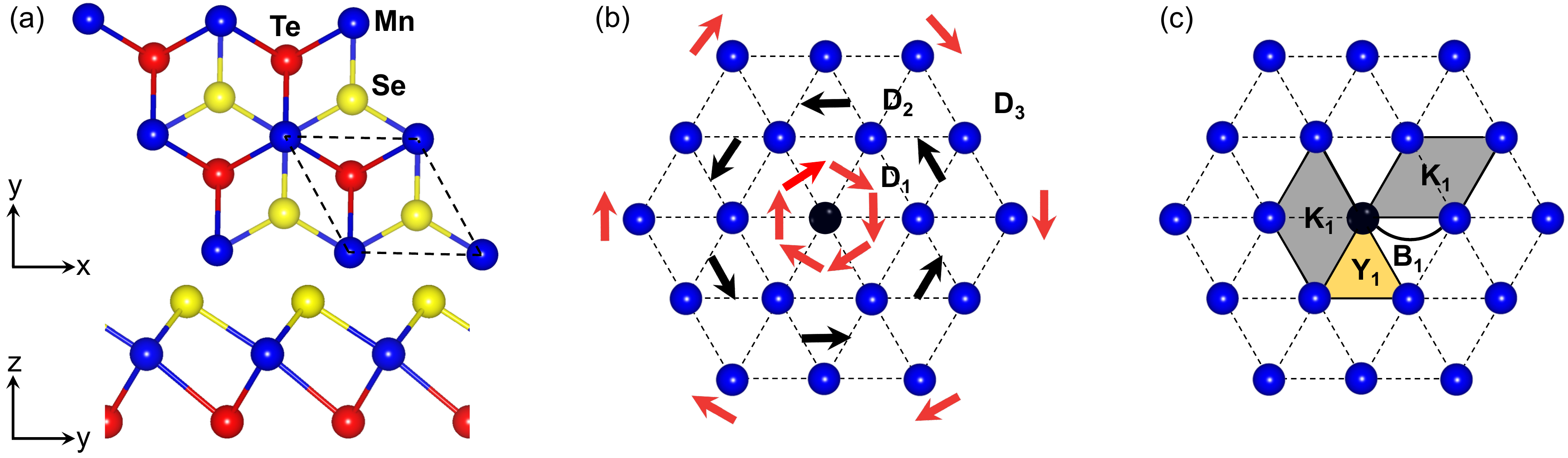}
    \caption{(a) Top (upper) and side (lower) views of the crystal structure of Janus MnSeTe. The black dashed lines indicate the 2D primitive cell.
(b) Illustration of DMI vectors on the hexagonal lattice. For clarity, only the Mn atoms are shown.
(c) Illustration of HOI constants on the hexagonal lattice. One of the six possible minimal hopping paths for the biquadratic interaction ($B_1$) is depicted by a black curve. One of the six possible minimal hopping triangular paths for the 4-spin 3-site interaction ($Y_1$) is depicted in yellow. Two of the twelve possible minimal hopping diamond-shaped paths for the 4-spin 4-site interaction ($K_1$) are depicted in gray.
}
    \label{Structure_Constants}
\end{figure*}

\begin{figure*}[t]%---------------------------------------------------------------------------------------------------------------------Fig_2
    \centering
    \includegraphics[width=0.85\linewidth]{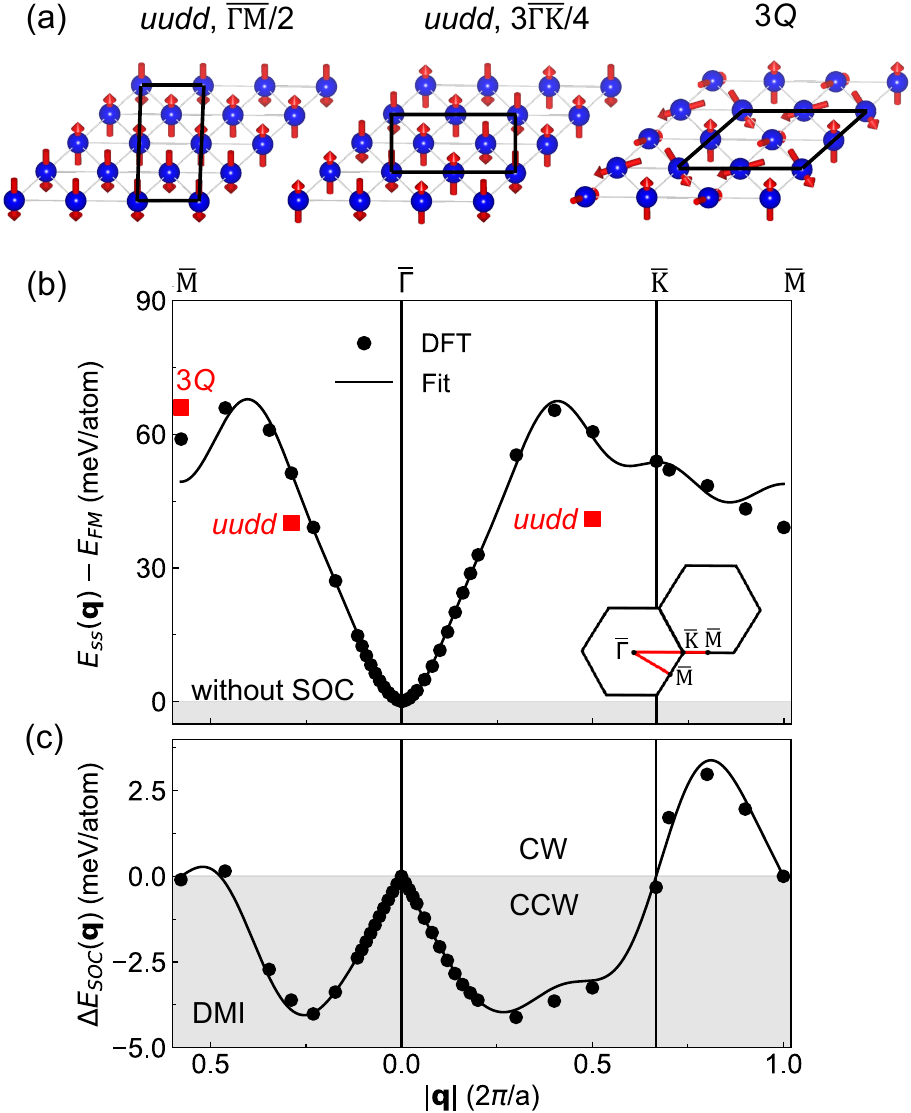}
    \caption{(a) Spin structures of multi-\textbf{q} states: two \textit{uudd} states along $\overline{\Gamma \text{M}}$ and $\overline{\Gamma \text{K}}$, and the $3Q$ state at the $\overline{\text{M}}$ point along $\overline{\Gamma \text{M}}$ of the 2D BZ.
(b) Energy dispersion of flat spin spirals ($E_{\text{ss}}$) for the MnSeTe monolayer along the high-symmetry path 
$\overline{\text{M} \Gamma \text{K} \text{M}}$
without SOC. The filled circles represent DFT total energies, while the solid lines are fits to the Heisenberg exchange interaction up to the tenth NN. The energies of the two \textit{uudd} and the $3Q$ states are also denoted by red squares at the \textbf{q} values of the corresponding single-\textbf{q} states.
(c) Energy contribution of flat spin spirals due to SOC ($\Delta E_{\text{SOC}}$), also referred to as the DMI contribution, is fitted up to the seventh NN. All energies are measured with respect to the FM state ($E_{\rm FM}$) at the $\overline{\Gamma}$ point. 
    }
    \label{SSstates_dispersion}
\end{figure*}%----------------------------------------------------------------------------------------------------------------------------

\begin{figure*}[t]%---------------------------------------------------------------------------------------------------------------------Fig_3
	\centering
	\includegraphics[width=0.6\linewidth]{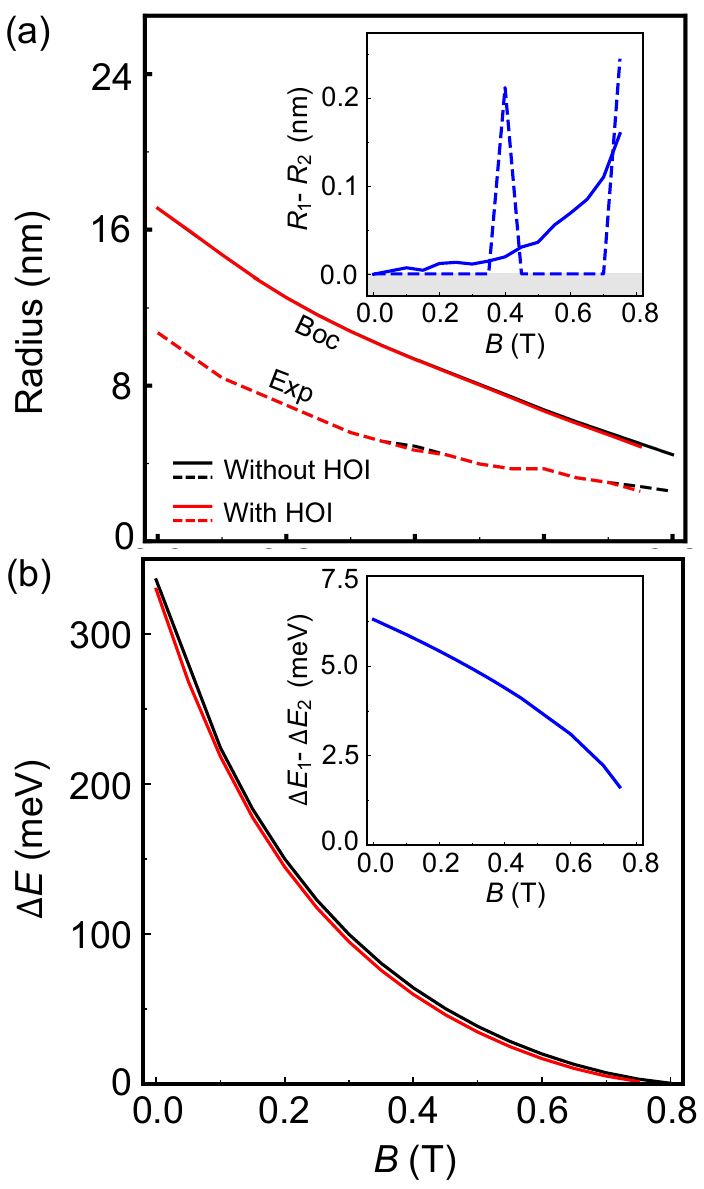}
	\caption{\label{radii_barrier} (a) Bogdanov (Boc) and experimental (Exp) skyrmion radii as functions of the applied magnetic field $B$ with (red) and without (black) HOI in Janus MnSeTe. The inset shows the radii difference with ($R_2$) and without ($R_1$) HOI, with the solid blue line for Boc and the dashed line for the Exp model. We observe that the Boc variation is smoother than the Exp one and hence numerically more stable. The inset further shows that the radii difference with and without HOI is very small, reaching maxima of about 0.15 nm at $B = 0.75$ T for the Boc model, and about 0.2 nm at $B = 0.4$ T and 0.22 nm at $B = 0.75$ T for the Exp model. (b) Total energy barriers of isolated skyrmions versus $B$ with and without HOI. The blue curve in the inset shows the difference between the two curves: black ($\Delta E_1$) and red ($\Delta E_2$).
    }
\end{figure*}%----------------------------------------------------------------------------------------------------------------------------

\begin{figure*}[t]%---------------------------------------------------------------------------------------------------------------------Fig_4
    \centering
    \includegraphics[width=1.0\linewidth]{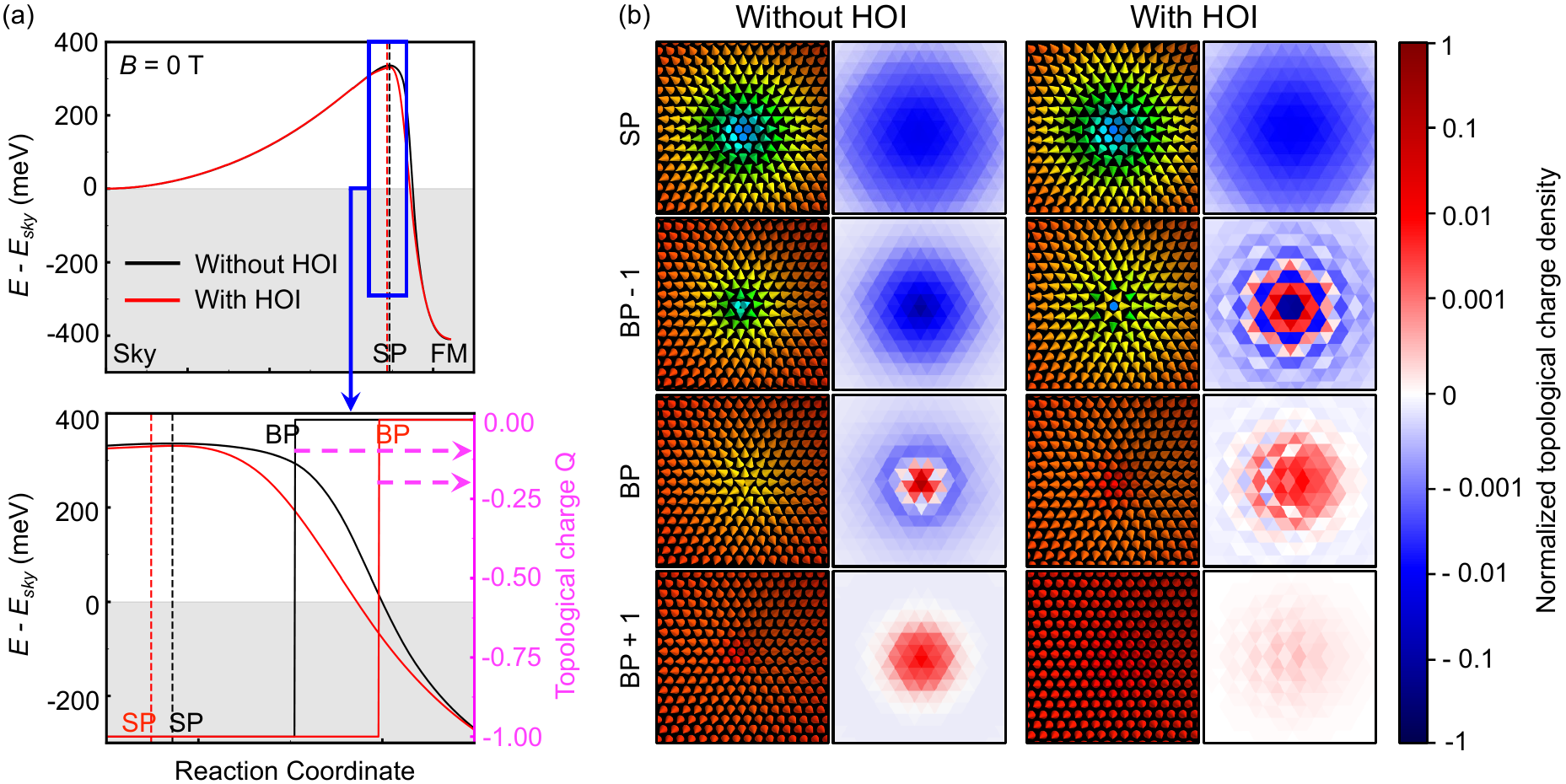}
    \caption{(a) MEP obtained for the transition from
    the skyrmion (sky) to the FM state through the SP for MnSeTe at $B = 0$ T with (red) and without (black) 
including HOI. The energies are shown w.r.t the skyrmion state. A zoom on the MEP near the SPs and the BPs is provided in the lower panel. SPs are marked as vertical dashed lines. BPs, identified via the magenta $y$-axis representing $Q$, are marked by vertical solid lines. (b) Corresponding spin textures (left panels) and topological charge densities (right panels) are also shown near SP and BP. Note that the topological phase transition occurs between BP-1 and BP. Interestingly, the skyrmion topological transition is radial without HOI, whereas it is ferric with HOI.
    }
    \label{MEP}
\end{figure*}%---------------------------------------------------------------------------------------------------------------------------

\begin{figure*}[t]%---------------------------------------------------------------------------------------------------------------------Fig_5
    \centering
    \includegraphics[width=1.0\linewidth]{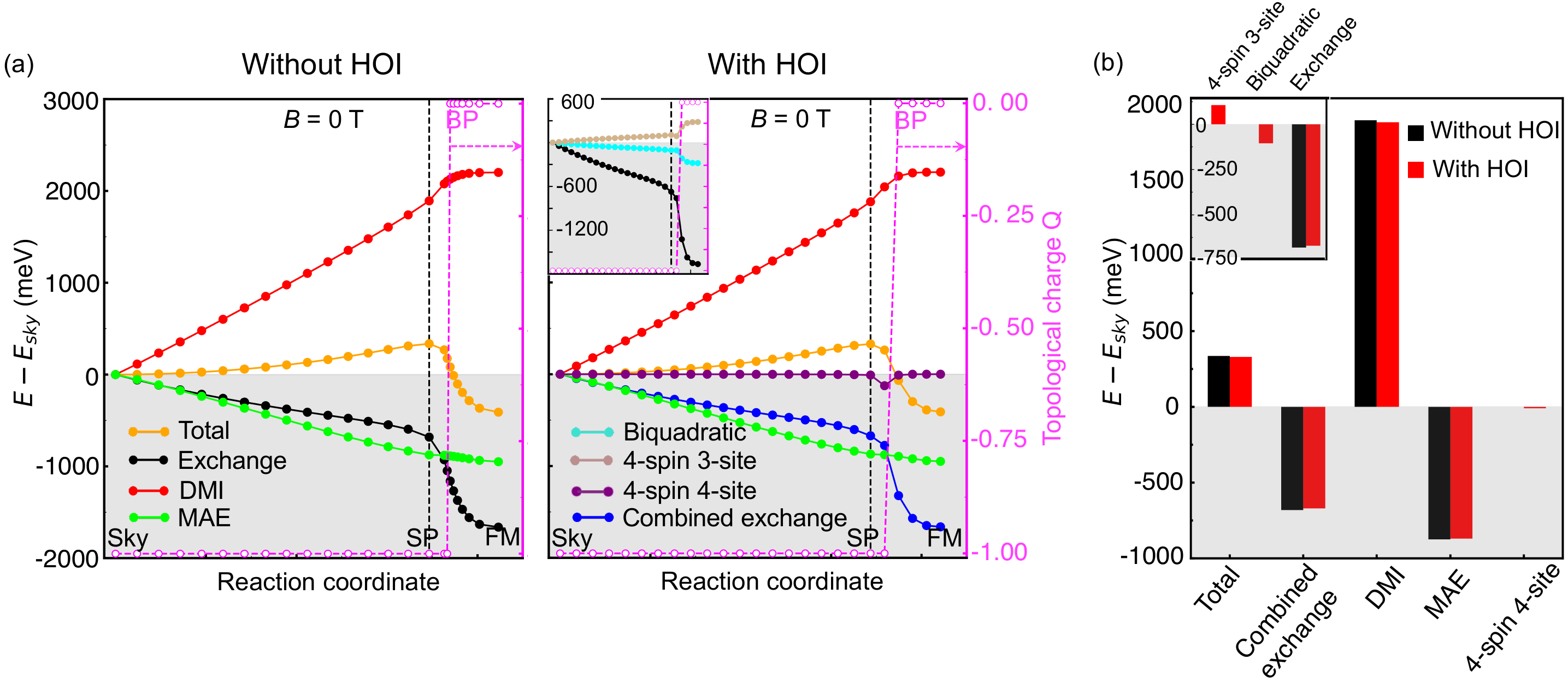}
    \caption{(a) Decomposition of the MEP obtained
    for the skyrmion to FM transition in MnSeTe
    with and without including HOI at $B = 0$ T. The energy contributions of the different interactions are shown w.r.t the skyrmion state energy $E_\text{sky}$ (see legend). On the right axis the topological charge along the MEP is shown (magenta curves). SP and BP are marked by the vertical dashed lines. (b) Energy decomposition at the SP as shown in (a) with respect to the skyrmion state, both with and without HOI. Note that combined exchange and exchange denote the same quantity without HOI.}
    \label{Decomposition}
\end{figure*}%--------------------------------------------------------------------------------------------------------------------------

\end{document}